\documentclass[%
 reprint,
nofootinbib,
nobibnotes,
 amsmath,amssymb,
 aps,
floatfix,
onecolumn
]{revtex4-2}

\usepackage{dcolumn}
\usepackage{bm}

\usepackage{amssymb,epsfig,latexsym}
\usepackage{amsmath,amscd}
\usepackage{graphicx,color}
\usepackage[normalem]{ulem}
\usepackage{subfigure}[ht]

\newcommand{\ijmpe}{Int. J. Mod. Phys. E.}
\newcommand{\apjs}{Astrophys. J. Suppl. }

\newcommand{\physrep}{Phys. Rep. }
\newcommand{\ptp}{Prog. Theor. Phys. }
\newcommand{\ptps}{Prog. Theor. Phys. Suppl. }
\newcommand{\ptep}{Prog. Theor. Exp. Phys. }

\newcommand{\aap}{Astron. Astrophys. }
\newcommand{\ssr}{Space Sci. Rev. }
\newcommand{\na}{Nat. Astron. }
\newcommand{\sci}{Science }

\newcommand{\mnras}{Mon. Not. Roy. Astron. Soc. }

\newcommand{\nphysa}{Nucl. Phys. A. }

\newcommand{\raa}{Res. Astron. Astrophys. }

\newcommand{\epl}{Europhys. Lett. }
\begin{document}

\title{Quiescent luminosities of accreting neutron stars with different equation of states}
\author{Helei Liu$^{1}$\footnote{email address: heleiliu@xju.edu.cn}
}
\author{Akira Dohi$^{2,3}$\footnote{email address: dohi@phys.kyushu-u.ac.jp}}
\author{Masa-aki Hashimoto$^{2}$}
\author{Yasuhide Matsuo$^{4}$}
\author{Guo-Liang L\"{u}$^1$} 
\author{Tsuneo Noda$^5$}
\affiliation{$^1$ School of Physical Science and Technology, Xinjiang University, Urumqi, $830046$, China
}
\affiliation{$^2$ Department of Physics, Kyushu University, Fukuoka, 819-035, Japan}
\affiliation{$^3$Interdisciplinary Theoretical and Mathematical Sciences Program (iTHEMS), RIKEN, Wako 351-0198, Japan}
\affiliation{$^4$ Nippo-cho, Minatokouhoku-ku, Yokohama, Kanagawa, 223-0057, Japan}
\affiliation{$^5$ Kurume Institute of Technology, Kurume, Fukuoka 830-0052, Japan}

\begin{abstract}
We model the quiescent luminosity of accreting neutron stars with several equation of states (EOSs), including the effect of pion condensation and superfluidity. As a consequence of comparison with the observations, we show that the results with Togashi EoS (the strong direct Urca process is forbidden) and TM1e EoS (mass at direct Urca process is $2.06 M_\odot$) can explain the observations well by considering pion condensation and the effect of superfluidity, while LS220 EoS and TM1 EoS can explain the observations well by considering the baryon direct Urca process and the effect of superfluidity. Besides, we compare the results with the observations of a neutron star RX J0812.4-3114 which has the low average mass accretion rate ($\langle\dot{M}\rangle\sim(4-15)\times 10^{-12}~M_\odot ~\rm yr^{-1}$) but high thermal luminosity ($L_q^\infty\sim(0.6-3)\times10^{33}~\rm erg ~ s^{-1}$), and we suggest that a low-mass neutron star ($<1M_\odot$) with minimum cooling can explain the lower limit of the observation of thermal luminosity of RX J0812.4-3114, which is qualitatively consistent with the previous work~\cite{Zhao2019}. However, to explain its upper limit, some other heating mechanisms besides standard deep crustal heating may be needed.
\end{abstract}

\maketitle

\section{Introduction}
It is known that thermal evolution of neutron stars (NSs) depends on the properties of superdense matter in NS cores, and thus it is of interest to attempt to study thermal states of transiently accreting NSs in the quiescent state~\cite{Yakovlev2003,Beznogov2015,Beznogov2,Han2017,Wijnands2017,Fortin2018,Potekhin2019,Kim2020,Liu2020}. By comparing the  theoretical curves (red-shifted quiescent luminosity $L_\gamma^\infty$ $-$average mass accretion rate $\langle\dot{M}\rangle$ diagram) with the available observationnal data of soft X-ray transients (SXRTs), we can explore the internal structure and equation of state (EoS) of dense matter.

SXRTs undergo periods of outburst activity (days to months, sometimes years) superimposed with longer periods (months to decades) of quiescence~\cite{Wijnands2017}. During an outburst, when the accretion is switched-on, the accreted matter is compressed under the weight of newly accreted material which leads to deep crustal heating with the release of $\thicksim1-2\rm~MeV$ per accreted matter~\cite{Haensel1990,Haensel2003,Haensel2008},
and the transient looks like a bright X-ray source ($L_X\sim10^{36}-10^{38}~\rm erg~s^{-1}$). During quiescence, since the accretion is switched-off or strongly suppressed, the NS luminosity decreases by several orders of magnitude ($L_X\lesssim10^{34}~\rm erg~s^{-1}$). The deep crustal heating is supposed to keep the NSs warm and explain the thermal emission of NSs in SXRTs~\cite{Brown1998}.

Great progress in observations of SXRTs in quiescence allowed us to pay attention to these objects.
 Yakovlev et al. were the first to examine the thermal state of transiently accreting NSs by using a simple toy model~\cite{Yakovlev2003};
Beznogov et al. studied the thermal evolution of NSs based on simulations of the evolution of stars of different masses and the presence of the powerful direct Urca (DU) neutrino emission process~\cite{Beznogov2015,Beznogov2};
Han et al. studied the thermal states of NSs in SXRTs by considering artificial DU process onset and neutrino emission~\cite{Han2017}; Fortin et al. studied the thermal states of NSs with a consistent model of interior~\cite{Fortin2018}, and they got the conclusion that the Brueckner-Hartree-Fock (BHF) EoS with the DU process is very well able to fit the current observations of isolated NSs and SXRTs. Potekhin et al. studied the thermal evolution and quiescent emission of transiently accreting NSs with the changes of the composition of the crust~\cite{Potekhin2019}.
In the above studies, they all feature strong DU processes, the kaon/pion condensations, hyperons and strange quark matter are not considered.
Matsuo et al. investigated the quiescent luminosities of accreting NSs with enhanced cooling due to pion condensation in the core of NSs~\cite{Matsuo2018}, but the effects of superfluidity were considered by artificial tuning of Gaussian distribution. Recently, an observational thermal luminosity of the Be/X-ray pulsar RX J0812.4-3114 has been reported~\cite{Zhao2019}, by comparing to theoretical predictions of the thermal luminosities produced by deep crustal heating for different time-averaged accretion rates, it shows that RX J0812.4-3114 lies above the minimal cooling tracks. As the system has a possibility of large magnetic field ($<8.4\times10^{11}~\rm G$) compared with typical low-mass X-ray binaries ($\sim10^9~\rm G$), it is interesting to study the high quiescent luminosity behavior of RX J0812.4-3114. 

On the other hand, the cooling of isolated NSs with different EoSs has been studied widely. For example, NS cooling with some microscopic BHF EoSs has been studied by Ref.~\cite{Wei2019}. NS cooling with some realistic EoSs (Togashi, Shen(TM1) and LS220 EoSs~\cite{Togashi2017,Shen2011,Lattimer1991NPA}) has been studied by Ref.~\cite{Dohi2019}, and they found that the NS cooling is slow with use of Togashi EoS due to its quite low symmetry energy, which is enough to prohibit the DU process with any masses. Meanwhile, the TM1 EoS always allows fast NS cooling, which can fit the observations by choosing suitable superfluid gap models (e.g. neutrons for $^{1}S_{0}$ state: CLS, 
for $^{3}P_{2}$ state: EEHO;
protons for $^{1}S_{0}$ state: CCDK in Ref.~\cite{Ho2015}). However, due to the large value of the symmetry energy slope $\simeq111~{\rm MeV}$, the radius is too large to be consistent with the constrain of the radius of NSs from GW170817 and the low mass X-ray binary observations which suggest a relatively small NS radii around 11-12 km with $M=1.4M_\odot$~\cite{Abbott2018,Steiner2010}. Recently, a new EoS based on the extended TM1 model has been constructed by Ref.~\cite{Shen2020}. The new EoS (TM1e) provides a similar maximum NS mass ($2.12M_\odot$) but smaller radius with a $1.4M_\odot$ star compared with TM1 model, which is more consistent with current constraints.
Since the TM1e EOS has lower symmetry energy slope $\simeq 40~{\rm MeV}$ than TM1 EOS, however, the threshold mass of DU process ($M_{DU}$) with TM1e EOS is $2.06M_{\odot}$ which is much higher than that with TM1 EOS. In a NS with $M<M_{DU}$, the DU process is turned off and the total neutrino emissivity is lower than for a NS with a mass above the $M_{DU}$. As a consequence, the former has a higher luminosity than the latter for a given age and accretion rate. Furthermore, the low luminosity observations of some SXRTs such as SAX J1808.4-3658 and 1H 1905+000 require fast cooling in the core of NS~\cite{Zhao2019,Heinke2009}.
Therefore, other fast cooling processes except the DU process, such as hyperon-DU process, quark-$\beta$ decay, and pion/kaon condensation, may work in NS. 
For instance, we can imply that the possibility of pion condensation in the core~\cite{Muto1993,Migdal1978,Maxwell1977,Tatsumi1983,Muto1987,Umeda1994}, the neutrino processes involving pions are much higher than the modified Urca process, thus the low temperature observations would be fitted.

According to the above investigations, the purpose of this work is to explore the quiescent luminosity of accreting NS with different EOSs: LS220, TM1, TM1e and Togashi . To explain the observations well with Togashi and TM1e EoSs, we construct the initial model with pion condensation. We also consider superfluid gap models and different surface composition. The paper is organized as follows. The approach to the modelling of thermal emission of SXRTs is described in Sec. \ref{sec:theory}, which includes the basic equations for NS thermal evolution and our physics inputs including EoSs (also EoSs with pion condensation), neutrino emission, superfluidity, deep crust heating and surface composition. In Sec. \ref{Res}, we show our results compared with the observations. After this, comparison of our models under minimal cooling with the thermal luminosity of RX J0812.4-3114 is discussed. Conclusions are drawn in Sec.\ref{Con}.

\section{Theory of thermal emission of SXRTs}\label{sec:theory}
In this work, we will model the quiescent phase of SXRTs. By using the spherically symmetric stellar evolutionary calculations firstly developed by Ref.~\cite{Fujimoto1984}, we then obtain the heating curves (relation between quiescent luminosity ($L_{\gamma}^{\infty}$) and time-averaged mass accretion rate $\langle\dot{M}\rangle$ ) of SXRTs. We will compute such heating curves for various EoSs which will turn on different neutrino emissivity. To fit the observations well, we will also consider different superfluid models and surface composition.

\subsection{Basic equations}
The full set of general relativistic structure and evolution equations can be written as follows~\cite{Thorne1977}:
\begin{equation}
    \frac{\partial M_{tr}}{\partial r}=4\pi r^2\rho,
 \label{eq:1}
\end{equation}

\begin{equation}
    \frac{\partial P}{\partial r}=-\frac{GM_{tr}\rho}{r^2}\left(1+\frac{P}{\rho \rm c^2}\right)\left(1+\frac{4\pi r^3P}{M_{tr}\rm c^2}\right)\left(1-\frac{2GM_{tr}}{r\rm c^2}\right)^{-1},
\label{eq:2}
\end{equation}

\begin{equation}
    \frac{\partial \left(L_{r}e^{2\phi/c^2}\right)}{\partial M_r}=e^{2\phi/c^2}\left(\varepsilon_n-\varepsilon_\nu\right),
\label{eq:3}
\end{equation}

\begin{equation}
   \frac{\partial {\rm ln} T}{\partial {\rm ln} P}=\frac{3}{16\pi acG}\frac{P}{T^{4}}\frac{\kappa L_{r}}{M_{r}}\frac{\rho_{0}}{\rho}\left(1+\frac{P}{\rho c^{2}}\right)^{-1}\left(1+\frac{4\pi r^{3}P}{M_{tr}c^{2}}\right)^{-1}\left(1-\frac{2GM_{tr}}{rc^{2}}\right)^{1/2}+\left(1-\left(1+\frac{P}{\rho c^{2}}\right)^{-1}\right),
\label{eq:4}
\end{equation}

\begin{equation}
    \frac{\partial M_{tr}}{\partial M_{r}}=\frac{\rho}{\rho_{0}}\left(1-\frac{2GM_{tr}}{r \rm c^2}\right)^{1/2},
\label{eq:5}
\end{equation}

\begin{equation}
    \frac{\partial \phi}{\partial M_{tr}}=\frac{G\left(M_{tr}+4\pi r^3 P/\rm c^2\right)}{4\pi r^4\rho}\left(1-\frac{2GM_{tr}}{r\rm c^2}\right)^{-1}.
\label{eq:6}
\end{equation}
where $M_{tr}$ and $M_r$ represent the gravitational and rest masses inside a sphere of radius $r$, respectively; $T$ and $P$ are the local temperature and the pressure, respectively; $\rho$ and $\rho_0$ are the total mass energy density and rest mass density, respectively; $\varepsilon _{n}$ denote the heating rate by nuclear burning, $\varepsilon_{\nu}$ is neutrino energy loss; $a$ is the Stefan-Boltzmann constant; $\kappa$ is the opacity and $\phi$ is the gravitational potential in unit mass. $G$ and $c$ are the gravitational constant and light velocity, respectively.

In the accretion layer, the mass fraction coordinate with changing mass ($q=M_r/M(t)$ where $M(t)$ is the total rest mass of the star) is adopted, which is the most suitable method for calculating stellar structure when the total stellar mass $M$ varies~\cite{Sugimoto1981}.
We set the outermost mass for numerical calculation as $q\sim10^{-20}$, which is sufficiently close to the photosphere. And then, we impose the radiative zero boundary condition expressed as follows~\cite{Fujimoto1984}:
\begin{eqnarray}
P &=& \frac{GMM(t)\left(1-q\right)}{4\pi R^4}\left(1-\frac{2GM}{Rc^2}\right)^{-1/2}~, \label{eq:A1} \\
L &=& \frac{4\pi cGM}{\kappa}\frac{4aT^4}{3P}\frac{1+\frac{\partial \log\kappa}{\partial \log P}}{4-\frac{\partial \log\kappa}{\partial \log T}}\left(1-\frac{2GM}{Rc^2}\right)^{1/2}~. \label{eq:A2}
\end{eqnarray}

Roughly speaking, these boundary conditions show that we view the luminosity at outermost mesh-point as the closest thing to the total luminosity $L$. Thus, we solve the set of equations by Henyey-type numerical scheme for the implicit method.


\subsection{Equation of state with pion condensation}

Under the assumption that NSs contain only neutrons, protons, electrons and muons, we adopt LS220, Togashi, TM1, and TM1e EoSs. The LS220 EoS is based upon the finite-temperature liquid-drop model with Skyrme-type interaction and the incompressibility is set to be $K=220~\rm MeV$~\cite{Lattimer1991NPA}, it is widely used in the study of supernovae and neutron stars~\cite{Couch2013,Couch2015,Oertel2017}. Togashi EoS is the first nuclear EoS based on realistic nuclear forces under finite temperature~\cite{Togashi2017}. TM1 EoS is based on the relativistic mean-field (RMF) theory. What is interesting, these three EoSs have been adopted to study the possibility of rapid NS cooling ~\cite{Dohi2019}.  TM1e EoS is based on an extended relativistic mean-field model with a smaller symmetry energy slope compared with TM1 EoS~\cite{Shen2020}. Based on the work of Ref.~\cite{Matsuo2018}, in which the quiescent luminosities of accreting NSs with LS220 EoS including pion condensation have been investigated. We then plan to include the exotic hadrons as pion condensation mixing in nuclear matter with the above four EoSs.
\begin{figure}[ht]
  \centering
  \includegraphics[width=0.8\linewidth]{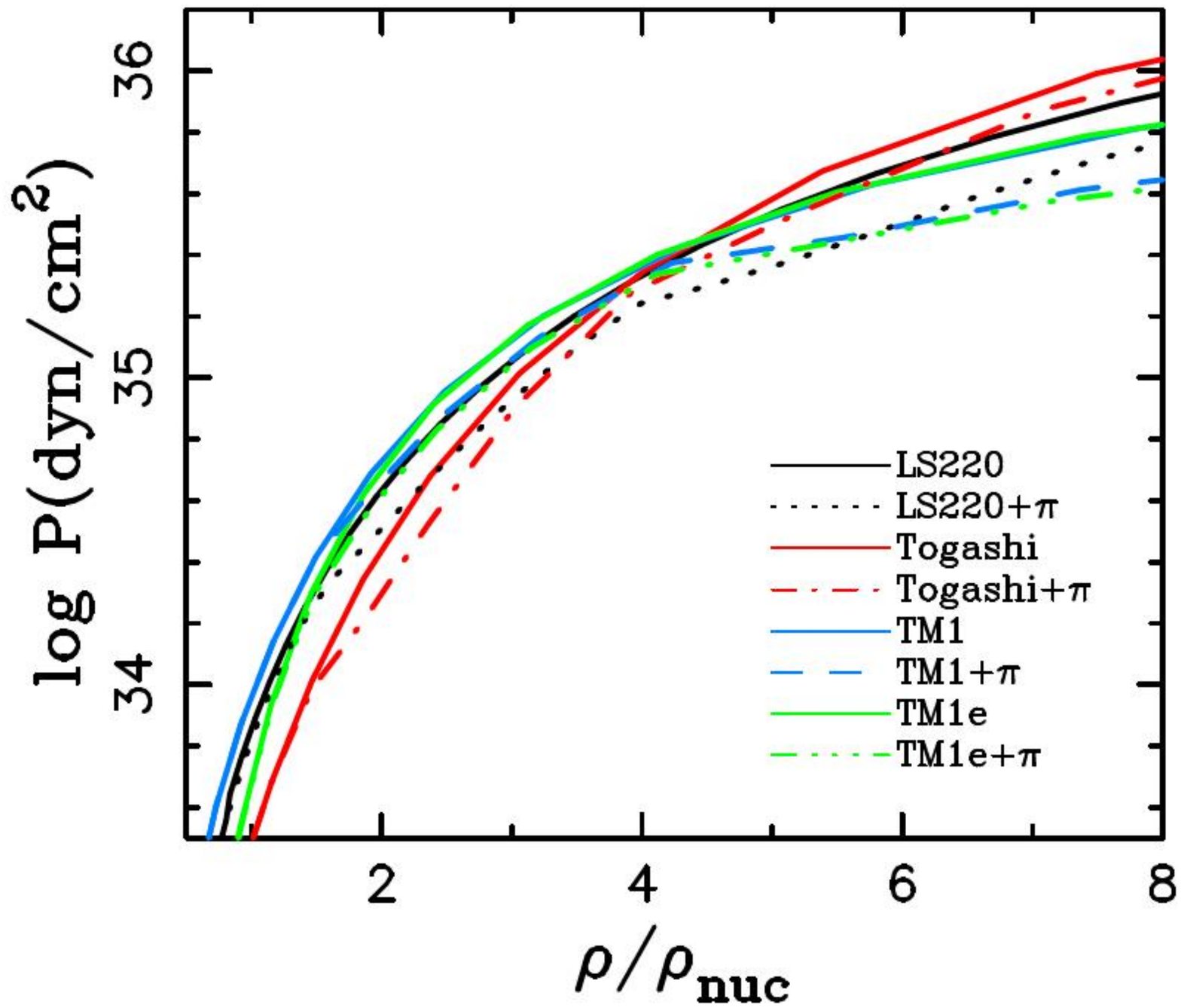}
  \caption{Pressure ($P$)-density ($\rho$) relation used in this work. The solid curves indicate EoS without pion condensation while dotted curves with pion condensation.}
  \label{fig:eos}
\end{figure}

\begin{figure}[ht]
  \centering
  \includegraphics[width=0.8\linewidth]{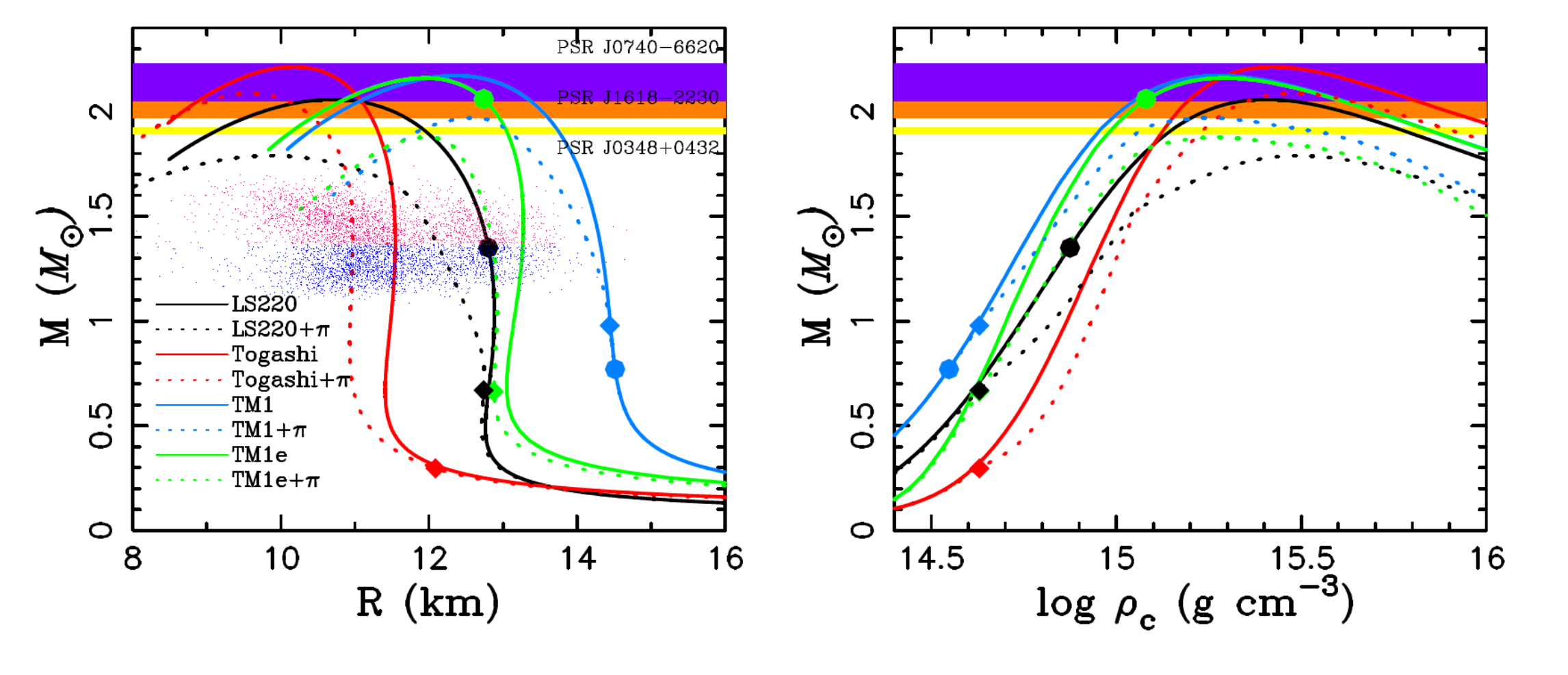}

  \caption{Left panel: Mass versus radius for the EoSs used in this work. Many red and blue dots indicate the results from the observation of GW170817~\cite{Abbott2018}; right panel: Mass versus central density for the EoSs used in this work. Marks on the curves show direct Urca thresholds. The shaded band shows the measurements of three pulsars: $2.14\pm0.09M_\odot$ for pulsar J0740-6620 (purple)~\cite{Cromartie2020}, $2.01\pm0.04M_\odot$ for pulsar J1618-2230 (orange)~\cite{Demorest2010}, $1.908\pm0.016$ for pulsar J0348+0432 (yellow)~\cite{Antoniadis2013}.}
  \label{fig:mr}
\end{figure}
The cooling of a NS with the pion condensation has been studied in Ref.~\cite{Umeda1994}. In the present work, we would construct the EoSs with the pion condensation firstly. The pion condensation affects NS cooling mainly in two ways: it softens the EoS at high densities, also, the enhancement of neutrino emissivity from pion Urca process would occur. We can obtain the EoS with pion condensation by adding pressure gain ($\Delta P$) and energy gain($\Delta \epsilon$) to the EoS, where $\Delta P$ and $\Delta\epsilon$ can be obtained from tables 1-2 of Ref.~\cite{Umeda1994}. In order to introduce strong pion Urca process in the cooling scenario to fit with the coldest transiently-accreting sources such as SAX J1808.4-3658~\cite{Han2017,Heinke2007}, we adopt the Landau-Migdal(LM) parameter $\tilde{g}'=0.5$ which corresponds to the strong pion Urca neutrino emissivity compared with $\tilde{g}'=0.6$~\cite{Umeda1994}. The EoSs with and without pion condensation are shown in Fig. \ref{fig:eos}. One can find that, by considering the effect of pion condensation, the EoS becomes softer. In Fig. \ref{fig:mr}, we present the mass-radius ($M-R$) and mass-central-density ($M-\rho_{c}$) curves for these EoSs. The shaded bands indicate the existence of three pulsars with measured mass around or above $2M_\odot$: pulsar J0740-6620~\cite{Cromartie2020}, pulsar J1618-2230~\cite{Demorest2010}, and pulsar J0348+0432~\cite{Antoniadis2013}. The measurements of these high mass pulsars ($\gtrsim 2M_\odot$) give a constrain on the EoS, for which the maximum mass should be around or above $2.0M_\odot$.
Many red and blue dots indicate the results from observation of GW170817~\cite{Abbott2018}. Obviously, from the mass-radius diagram, Togashi EoS is more suitable to fit these observational constraints than others. The solid curves indicate the $M-R$, $M-\rho_c$ relations without pion condensation while the dotted curves with pion condensation. It shows that the maximum mass of NS with LS220 EoS is reduced from $2.06M_\odot$ to $1.79M_\odot$ due to pion condensation, while for Togashi EoS, the maximum mass is reduced from $2.21M_\odot$ to $2.09M_\odot$; for TM1 EoS, the maximum mass is reduced from $2.17M_\odot$ to $1.97M_\odot$; for TM1e EoS, the maximum mass is reduced from $2.16M_\odot$ to $1.88M_\odot$. Selecting EoSs with the mass constrain from the above three pulsars, we will exclude LS220+$\pi$ EoS in the following quiescent luminosity calculations. The dots on the solid curves show nucleon DU thresholds ($M_{DU}$), $M_{DU}=1.35M_\odot$ for LS220 EoS, $M_{DU}=0.77M_\odot$ for TM1 EoS, $M_{DU}=2.06M_\odot$ for TM1e EoS. As the DU process is forbidden for Togashi EoS, we don't make mark on the curve with Togashi EoS. The rhombuses on the dashed curves show pion DU thresholds, the critical density where the pion phase appears is $1.67\times10^{15}\rm ~g ~cm^{-3}$, the corresponding pion Urca thresholds are $0.67M_\odot$ for LS220+$\pi$ EoS, $0.30M_\odot$ for Togashi+$\pi$ EoS, $0.98M_\odot$ for TM1+$\pi$ EoS, $0.66M_\odot$ for TM1e+$\pi$ EoS. One can see $M_{DU}$ clearly from Table \ref{tab:DU}.

\begin{table}[tbp]
\caption{Direct Urca or Pion Urca thresholds for the EoSs used in this work. }
\label{tab:DU}
\centering
\begin{tabular}{ccccccccc}
\hline\hline
EoS & LS220 & Togashi & TM1 & TM1e & LS220+$\pi$ & Togashi+$\pi$ & TM1+$\pi$ & TM1e+$\pi$ \\
\hline
$M_{DU}/M_\odot$ & 1.35 & - & 0.77 & 2.06 & 0.67 & 0.30 & 0.98 & 0.66\\
\hline
\hline
\end{tabular}
\end{table}

\subsection{Neutrino emission mechanisms and superfluidity}

One of the key ingredients of quiescent luminosity simulation is the neutrino cooling processes. In a NS containing only neutrons, protons, electrons and muons, the most powerful neutrino emission process is the baryon DU process:
\begin{equation}
    n\rightarrow p+e^{-}+\bar{\nu}_e, \qquad   {\rm and} \qquad   p+e^{-}\rightarrow n+\nu_e,
 \label{eq:DU}
\end{equation}
For this process, the neutrino emissivity is about $10^{27}T_9^6~\rm erg~cm^{-3}~s^{-1}$(see Eq.(120) in Ref.~\cite{Yakovlev2001} for detail), where $T_9$ is the local temperature in units of $10^9~\rm K$. However, the energy and momentum conservation imposes a threshold of proton fraction of whether to cause this process~\cite{Lattimer1991}, where the critical proton fraction $Y_p=1/9$ if $Y_\mu=0$, and thus the DU process operates in the central part of NS with masses larger than $M_{DU}$, corresponding to the dots on the curves of Fig. \ref{fig:mr}.

For a NS with the central pion-condensation core, the strongest reactions are the $\eta$-particle Urca process:
\begin{equation}
\begin{aligned}
 \eta(\textbf{p})  \rightarrow \eta(\textbf{p}')+e^{-}(\textbf{p}_e)+\bar{\nu}_e(\textbf{p}_\nu),\\
 \eta(\textbf{p})+e^{-}(\textbf{p}_e)  \rightarrow \eta(\textbf{p}')+\nu_e(\textbf{p}_\nu)
\end{aligned}
 \label{eq:pionDU}
\end{equation}
The neutrino emission rates due to pion condensation is about $10^{25}T_9^6~\rm ergs ~cm^{-3} ~s^{-1}$(see tables 1-4 in Ref.~\cite{Umeda1994} for detail).

If the DU process is forbidden or strongly reduced, two less efficient neutrino processes are operated mainly by the modified Urca (MU) process:
\begin{equation}
    n+N\rightarrow p+e^{-}+\bar{\nu}_e+N \qquad   {\rm and} \qquad p+e^{-}+N\rightarrow n+\nu_e+N
 \label{eq:MU}
\end{equation}
where $N$ is a spectator nucleon that ensures momentum conservation,
and nucleon-nucleon bremsstrahlung process:
\begin{equation}
  N+N\rightarrow N+N+\nu+\bar{\nu}
 \label{eq:bre}
\end{equation}
where $N$ is a nucleon, these two neutrino emission rates are approximately $10^{19-21}T_9^8~\rm ergs ~cm^{-3} ~s^{-1}$(see Eqs.(140) and (142) in Ref.~\cite{Yakovlev2001} for detail).
Besides, the neutrino emission due to electron-ion, electron-positron pair, photon and plasmon processes are also included in this work~\cite{Yakovlev2001}. It is worth noting that more accurate expressions for the electron-ion neutrino bremsstrahlung have been derived~\cite{Ofengeim2014}. 

\begin{figure}[ht]
  \centering
  \includegraphics[width=0.8\linewidth]{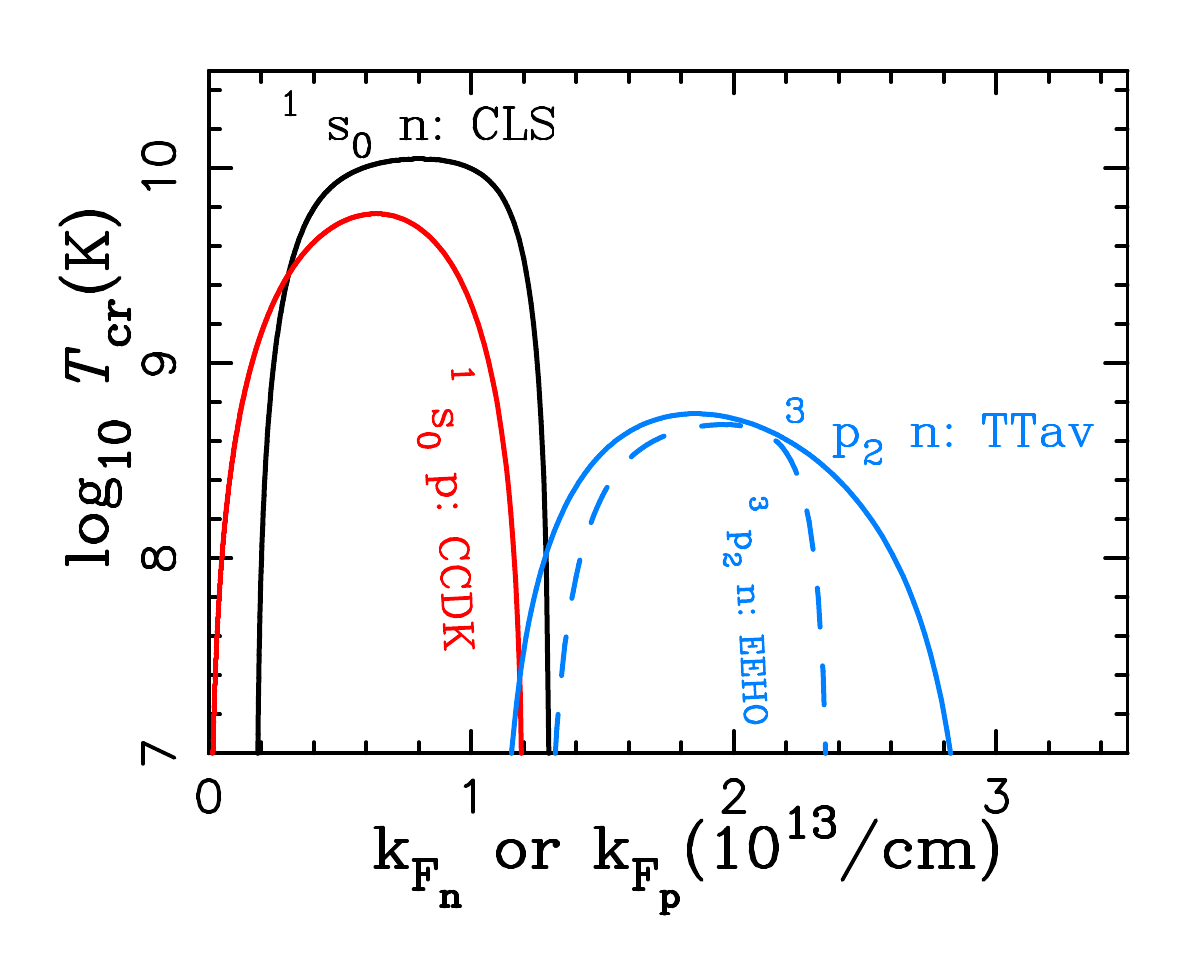}
  \caption{Critical temperature $T_{cr}$ as a function of fermi momentum. Different curves represent different superfluid models, which come from Ref. ~\cite{Ho2015}.}
  \label{fig:sf}
\end{figure}

The existence of superfluidity in NSs had been confirmed long before by the theory and observations~\cite{Yakovlev2005}.
The effect of superfluidity on the neutrino emissivity is twofold: one is to reduce the neutrino emission rate exponentially when the temperature decreases below the critical superfluid temperature $T_{\rm cr}$, another is to enhance neutrino emission process due to pair breaking and formation (PBF) when the temperature decreases just below $T_{\rm cr}$~\cite{Lattimer2004}, we adopt expressions of PBF process from Ref.~\cite{Lattimer2004} in this work. However, one should note that the expressions of PBF process were revised by taking into account of anomalous weak interactions~\cite{Leinson2009,Leinson2010}, which would reduce the neutrino emissivity by a factor of $\sim0.2$ compared with Ref.~\cite{Lattimer2004}. Neutron $^{1}S_{0}$, proton $^{1}S_{0}$, neutron $^{3}P_2$ channels on the neutrino emissivity are considered, and we adopt the critical temperature the same as those used in Ref.~\cite{Ho2015}. Their corresponding $T_{\rm cr}(k_F)$ curves are displayed in Fig.~\ref{fig:sf}, where $k_F$ is the Fermi momentum.

\subsection{Deep crustal heating and envelope composition}
The heating rate $\varepsilon_n$ in eq.~(\ref{eq:3}) is generated by deep crustal heating in the present work. The deep crustal heating has the following form~\cite{Haensel1990}:
\begin{equation}
Q_i=6.03\times\dot{M}q_i 10^{43} ~\rm erg~s^{-1},
 \label{eq:heating}
\end{equation}
where $q_i$ is the effective heat per nucleon on the $i$-th reaction surface. We adopt Ref.~\cite{Haensel2008} as the heating rate where the initial compositions of the nuclear burning ashes are fixed to be ${}^{56}{\rm Fe}$ (for detail, see table A.3 reference therein).

The envelope with light elements has higher surface temperature and thus surface photon luminosity at neutrino cooling stage. Hence we include two extreme cases: pure Ni envelope and pure He envelope with $\Delta M/M_{\rm NS}=10^{-7}$~\cite{Potekhin1997}, where $\Delta M$ is the envelope mass and $M_{\rm NS}$ is the gravitational mass. For the opacity $\kappa$ in Eq.~(\ref{eq:4}), we take into account radiative opacity~\cite{Schatz1999} and the conductive opacity which is composed of mainly electrons~\cite{Potekhin2015} and neutrons~\cite{Baiko2001}.


\section{Results and discussion}\label{Res}
\subsection{The quiescent luminosity of accreting neutron star with different EoSs }
One important tool of analysis is the quiescent luminosity($L_\gamma^\infty$) vs. time-averaged mass accretion rate ($\langle\dot{M}\rangle$) diagram, in which currently a few ($\sim35$) observed accreting NSs are located~\cite{Potekhin2019}. Using the specified EoSs in Fig.~\ref{fig:eos} except LS220+$\pi$  EoS for which the maximum mass is less than $2.0M_\odot$ and TM1+$\pi$ EoS as the TM1 EoS operate DU process at any mass and we don't need to include other fast cooling process such as pion condensation, the thermal evolution of accreting NSs are calculated.

\begin{figure}[ht]
  \centering\hspace*{-2cm}
  \includegraphics[width=0.6\linewidth]{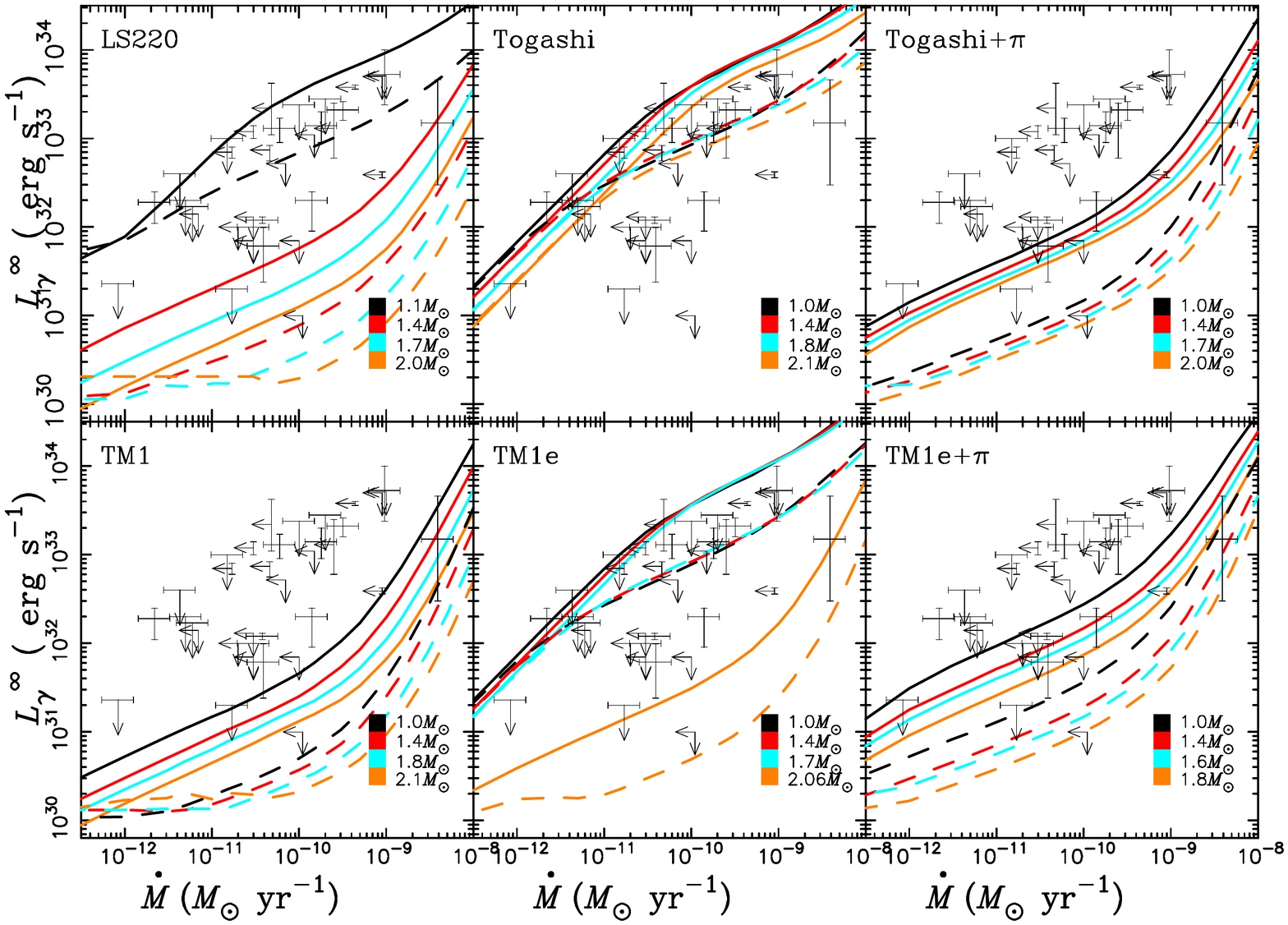}
  \caption{Quiescent luminosities of SXRTs as functions of time-average mass accretion rates, without nucleon superfluidity. Different panels indicate the models with different EoSs. The solid curves indicate the pure He envelope ($\Delta M/M=10^{-7}$) while the dashed curves indicate the pure Ni envelope. Different masses are marked by color. The errorbars in each panel are taken from Table 2 of Ref.~\cite{Potekhin2019}.}.
  \label{fig:qlnosf}
\end{figure}

Fig.~\ref{fig:qlnosf} shows the redshifted quiescent luminosities of NSs in SXRTs as a function of time-averaged mass accretion rate.
 For LS220 EoS, since the DU threshold operate at $M_{DU}\simeq1.35M_\odot$, the curves locate too low in the panel with $M>1.4M_\odot$ that the observations can't be explained well. We need to include the effect of superfluidity to suppress the fast cooling due to DU process. We see that Togashi EoS is unable to explain the whole range of the estimated values of $L_q^\infty$ and $\langle\dot{M}\rangle$ simultaneously. Because the DU process is forbidden for Togashi EoS, the quiescent luminosities are high with a fixed mass accretion rate. It is necessary to include other fast cooling in the core of Togashi EoS. For Togashi+$\pi$ EoS, as the pion DU threshold is $0.3M_\odot$, the NSs cool fast that the curves with $M\geq1.0M_\odot$ are located in the lower panel, in order to explain the observations well, we need include the effect of superfluidity. The case of TM1 EoS is similar with Togashi+$\pi$ EoS, as the DU process is operate at $0.77M_\odot$, the NSs cool too fast that the curves can't explain the observations even with $1.0M_\odot$  NS. While for TM1e EoS, the DU process is operate at $M_{DU}=2.06M_\odot$, we also need include other fast cooling as pion condensation to fit the observations as seen in the middle of bottom panel of Fig.~\ref{fig:qlnosf}. For TM1e+$\pi$ EoS, as the pion DU threshold is $0.66M_\odot$, the results are unable to explain the observations with high luminosities, which indicates that the superfluidity is also needed for TM1e+$\pi$ EoS in order to suppress the fast cooling due to pion condensation.

\begin{figure}[ht]
\vspace*{-1.5cm}
  \centering\hspace*{-1.0cm}
  \includegraphics[width=0.6\linewidth]{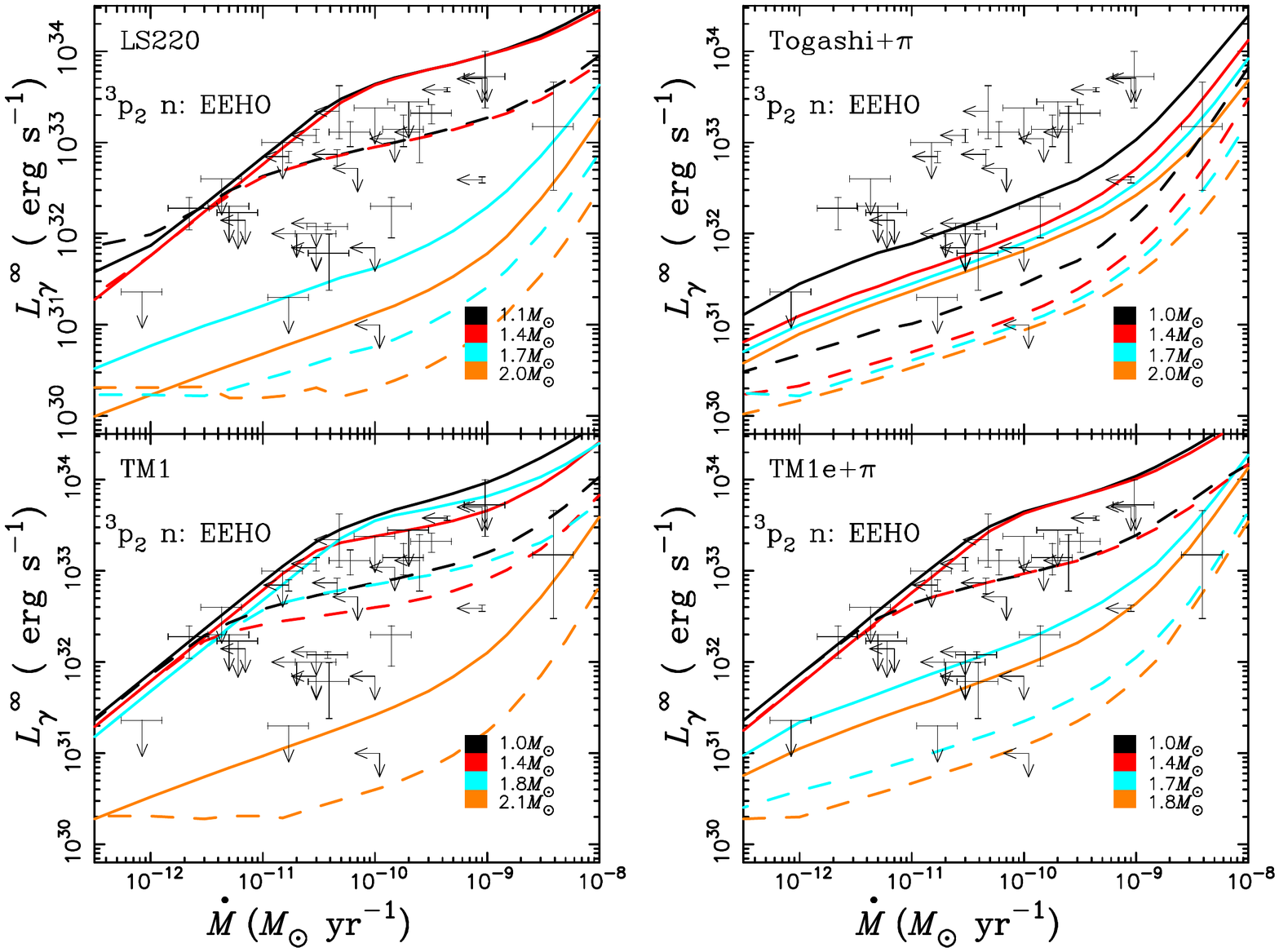}
   \caption{Quiescent luminosities of SXRTs as functions of time-average mass accretion rates, but considering the effect of nucleon superfluidity. The superfluidity models are as follows: CLS for neutron $^{1}S_0$, CCDK for proton $^{1}S_0$, EEHO for neutron $^{3}P_2$, the critical temperature for the models are shown in Fig.~\ref{fig:sf}.}
  \label{fig:EEHO}
  
   \centering\hspace*{-1.0cm}
  \includegraphics[width=0.6\linewidth]{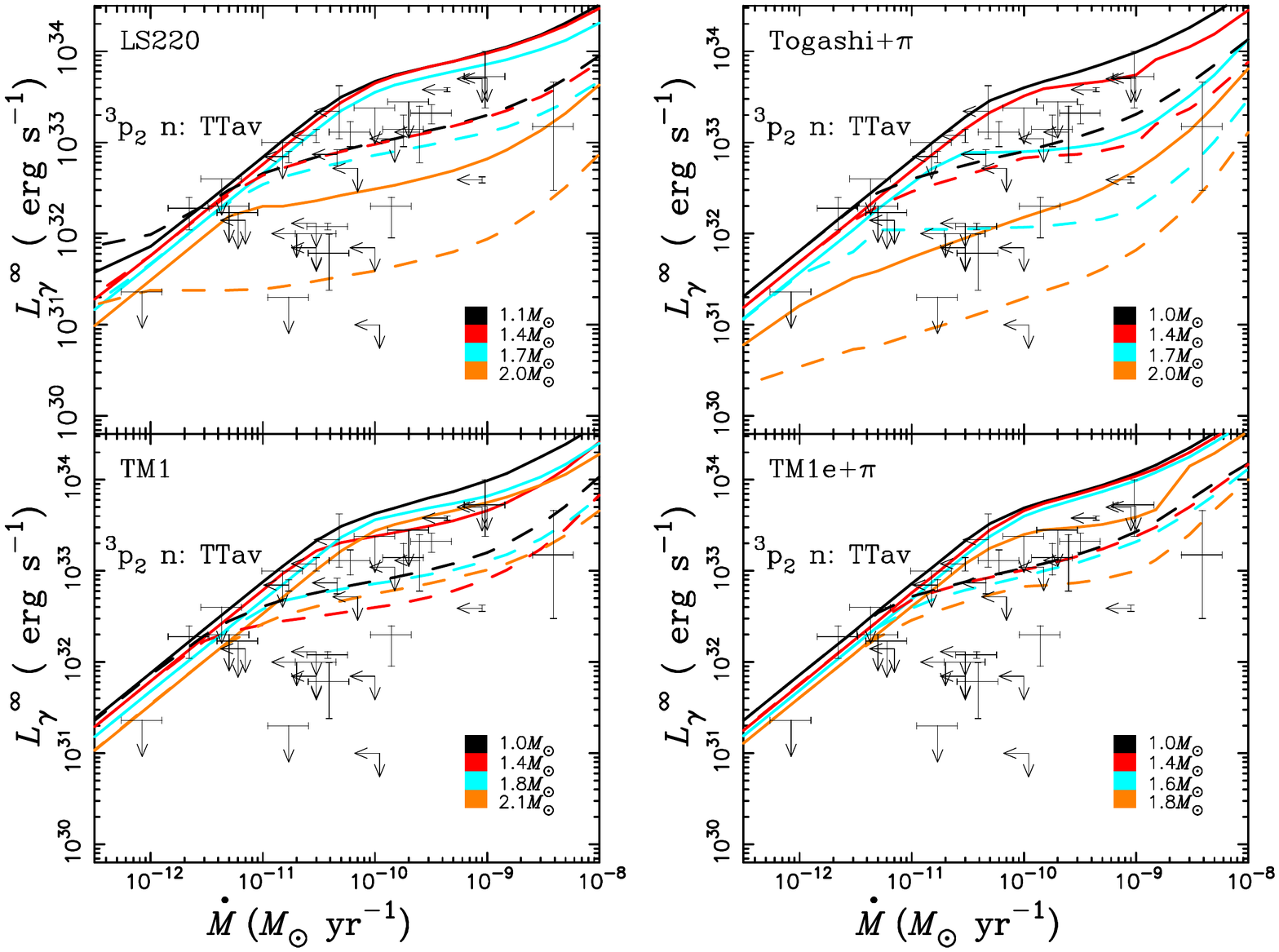}
  \caption{The same as Fig.~\ref{fig:EEHO} but with different neutron $^{3}P_2$ superfluidity model as TTav.}
  \label{fig:TTav}
\end{figure}

In Fig.~\ref{fig:EEHO}, we examine the effects of superfluidity on the $L_\gamma^\infty-\dot{M}$ curves. The superfluidity models of CLS for neutron $^{1}S_0$, CCDK for proton $^{1}S_0$, EEHO for neutron $^{3}P_2$ are adopted, the critical temperature of the models can be found in Fig.~\ref{fig:sf}. Here the thermal evolution of accreting NSs using the same model as in Fig.~\ref{fig:qlnosf} but with the effect of superfluidity. As the quiescent luminosities with Togashi and TM1e EoSs are high enough as shown in Fig.~\ref{fig:qlnosf}, we don't include these two EoSs in Fig.~\ref{fig:EEHO}.
We note that the observations could potentially be explained by regarding the effect of superfluidity. However, the effect of superfluidity adopted in the figure seems a little weak for LS220 and Togashi+$\pi$ EoS, as one can see from Fig.~\ref{fig:EEHO}, for LS220 EoS, the location of the curves with $M\geq1.7M_\odot$ change a little compared with that in Fig.~\ref{fig:qlnosf}, while for Togashi+$\pi$ EoS, only the curves with $1.0M_\odot$ change significantly to fit the high luminosities observations, for $M\geq1.4M_\odot$, the curves don't change obviously compared with that in Fig.~\ref{fig:qlnosf}. Due to this problem, Fig.~\ref{fig:TTav} shows the curves with more strong neutron $^{3}P_2$ model as TTav. As can be seen in Fig.~\ref{fig:sf}, TTav has wider superfluid effect in high density region than EEHO model, as a result, the former would have stronger effect than latter, the curves in Fig.~\ref{fig:TTav} are enhanced compared with Fig.~\ref{fig:EEHO}. For LS220 and Togashi+$\pi$ EoSs, most of the observations can be fitted well except the coldest one, but the suppression of $^{3}P_2$ model is too strong for TM1 and TM1e+$\pi$ EoSs, the quiescent luminosity is too high for a fixed accretion rate.

\begin{figure}[ht]
  \centering\hspace*{-1.0cm}
  \includegraphics[width=0.9\linewidth]{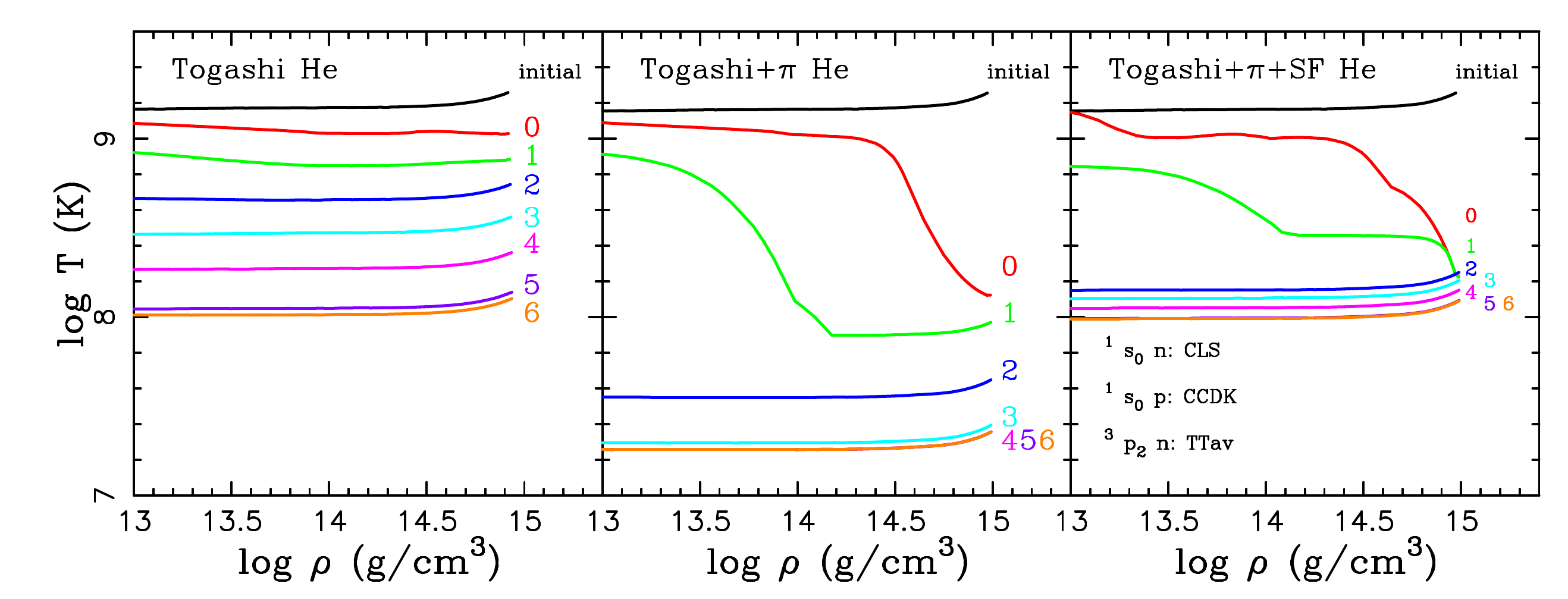}
  \caption{Time evolution of local temperature towards the steady state of NSs. The left and middle panels show the results for Togashi, Togashi+$\pi$ EoSs with $1.4M_\odot$, respectively. The right panel is same as middle panel but include the effect of superfludity, which we marked as Togashi+$\pi$+SF. The accretion rate is set as $1\times10^{-10} \rm~M_\odot~ yr^{-1}$ and the surface composition is pure He for the calculations. The numerals attached indicate the ages of log$~t~\rm{(yr)}$.}
  \label{fig:stru}
\end{figure}

In the models described above, large changes of the quiescent luminosities result from adopting different EoSs which determine the fast cooling process and superfluidity models which suppress the too strong cooling process. In Fig.~\ref{fig:stru}, we show the temperature as a function of density for Togashi EoS in three cases as an example. Without superfluidity, the thermal structure of the models with the Togashi EoS ($1.4M_\odot$) show that the DU process doesn't operate in the core of the star, this case correspond to the slow cooling, so the heating curves with Togashi EoS are unable to explain the whole range of the observations in Fig.~\ref{fig:qlnosf}.  While for Togashi+$\pi$ EoS in the middle panel, the pion DU process works in the core of the star, and the temperature of the core drop rapidly at the first 0-100 years which results in the low temperature at steady state. As a result, the $L_\gamma^\infty-\dot{M}$ curves with Togashi+$\pi$ EoS locate too low to fit the observations. In the right panel of Fig.~\ref{fig:stru} shows the effect of superfluidity on the thermal structure of accreting NS with Togashi+$\pi$ EoS, the rapid temperature drop is suppressed by the effect of superfluidity compared with the middle panel, so the observations can be explained well with Togashi+$\pi$ EoS in Fig.~\ref{fig:TTav}.

From Figs.~\ref{fig:qlnosf}, \ref{fig:EEHO}~and~\ref{fig:TTav}, we conclude that for LS220 and TM1 EoSs, which have low DU threshold, one can fit the observations well with those EoSs by considering proper superfluidity models besides different envelope composition and a range of masses. While for the EoSs such as Togashi and TM1e, for which the DU threshold is too high or forbidden, we can include the other fast cooling process such as pion condensation in the core of NS to operate fast cooling process, the models with Togashi+$\pi$ and TM1e+$\pi$ EoSs can also explain the observations well by choosing proper superfluid models.

\subsection{Thermal luminosity of RX J0812.4-3114 with minimal cooling}
\begin{figure}[ht]
  \centering\hspace*{-1.0cm}
  \includegraphics[width=0.6\linewidth]{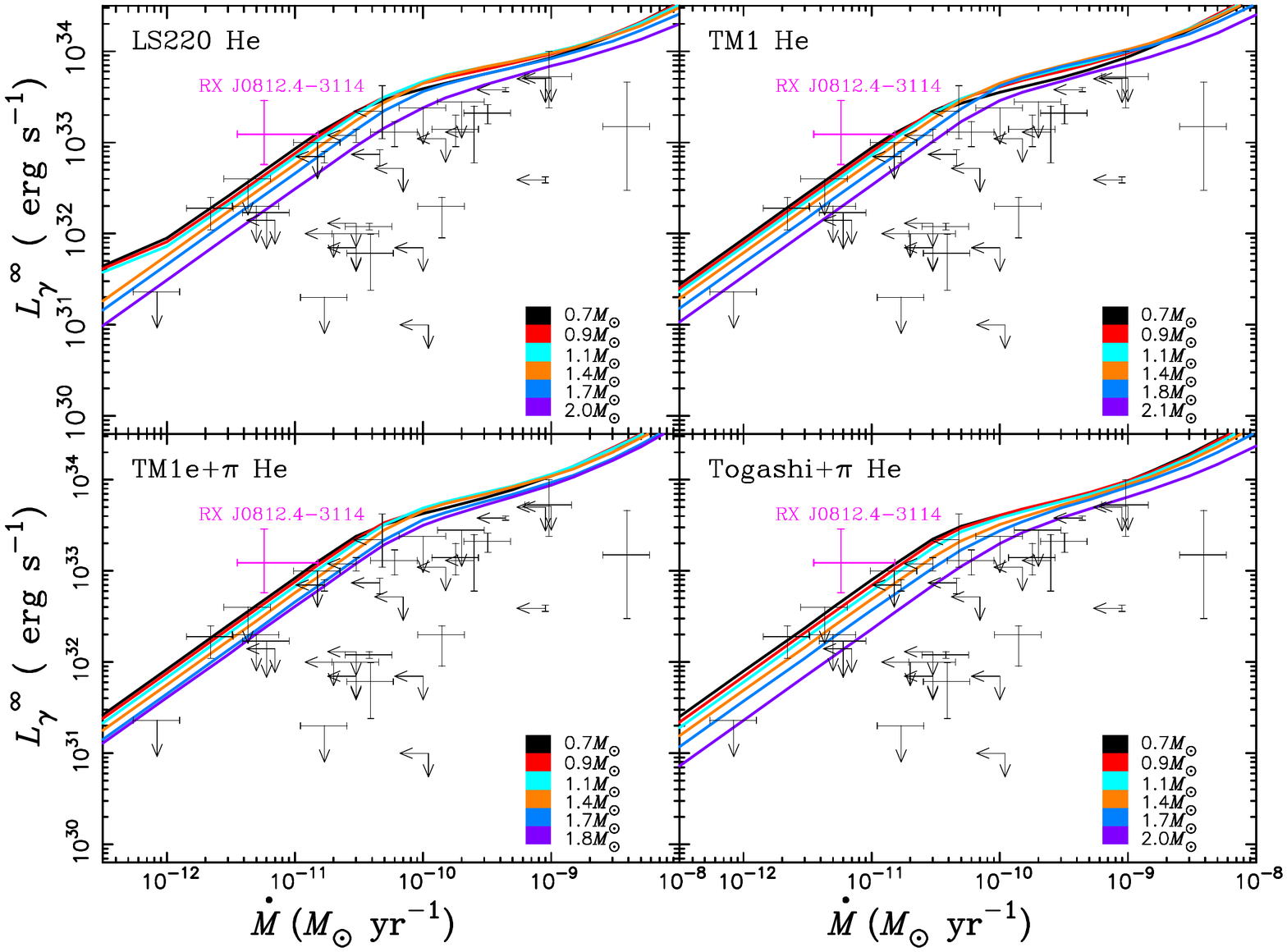}
  \caption{The same as Fig.~\ref{fig:EEHO} but with minimal cooling for which the DU process are turned off. The pink error indicate the observation of RX J0812.4-3114.}
  \label{fig:mini}
\end{figure}

The quiescent luminosity of Be/X-ray pulsar (BeXRP) RX J0812.4-3114 has been estimated as $L_q^\infty\sim(0.6-3)\times10^{33}~\rm erg ~ s^{-1}$, and its time-average mass accretion rate is estimated as $\langle\dot{M}\rangle\sim(4-15)\times10^{-12}~\rm{M_\odot}~ yr^{-1}$~\cite{Zhao2019}. It has been shown that the thermal luminosity of RX J0812.4-3114 is too high to be explained by the standard deep crustal heating model.
 There are two possible explanations: RX J0812.4-3114 may contain a low-mass NS with minimum cooling, or the system may be young enough that the NS still hot from supernova explosion~\cite{Zhao2019}. We verify the former assumption based on our work. In minimal cooling scenario, the fast cooling from any DU process will not be included. So we turn off baryon DU process for LS220 and TM1 EoSs, and pion DU process for Togashi+$\pi$ and TM1e+$\pi$ EoSs. For the effect of superfluidity, we choose the same model as used in Fig.~\ref{fig:EEHO}: CLS for neutron $^{1}S_0$, CCDK for proton $^{1}S_0$ and EEHO for neutron $^{3}P_2$. The results can be found in Fig.~\ref{fig:mini}, it is shown that the minimal cooling with small mass NS ($<1M_\odot$) can probably fit the lower limit of the high thermal luminosity of RX J0812.4-3114, no matter for LS220, TM1, TM1e+$\pi$ or Togashi+$\pi$ EoSs. Our results qualitatively agree with Ref.~\cite{Zhao2019}, while the quantitative differences may be caused by the different microphysics input. The upper limit luminosity of RX J0812.4-3114 can't be fitted by the standard deep crustal heating model, which indicates that the NS in RX J0812.4-3114 is too hot. One possible way to explain the upper limit luminosity of RX J0812.4-3114 is to consider that it is still hot from supernova explosion as the previous work mentioned~\cite{Zhao2019}. Another possible way is that there are other heating mechanisms in addition to standard deep crustal heating in RX J0812.4-3114. We need further observations to understand more about the physics in RX J0812.4-3114.

\section{Conclusions}\label{Con}
Motivated by the cooling of NS is slow with Togashi EoS, the DU threshold is high for TM1e EoSs and the recent availability of more stringent restrictions on the EoSs of NS provided by GW170817. We have computed the quiescent luminosities of accreting NSs in this work with different EoSs (LS220, TM1, TM1e, Togashi, Togashi+$\pi$ and TM1e+$\pi$) by using stellar evolutionary calculations. As the DU threshold is low for LS220 ($M_{DU}\approx1.35M_\odot$) and TM1 ($M_{DU}\approx0.77M_\odot$) EoSs, we can simulate the quiescent luminosity with those two EoSs to fit the observations well by considering the effect of superfluidity besides the different surface composition and different masses. However, for Togashi and TM1e EoSs, their DU threshold is forbidden or too high, as a result, the steady luminosity is too high with these two EoSs compared with LS220 and TM1 EoSs. To fit the observations well, we include pion condensation with them, named as Togashi+$\pi$ and TM1e+$\pi$, respectively. As the pion DU threshold is $0.3M_\odot$ for Togashi+$\pi$ EoS and $0.66M_\odot$ for TM1e+$\pi$ EoS, the simulations of quiescent luminosities with those two EoSs are low and we can improve them also by choosing a proper superfluid model. Besides, the thermal luminosity of RX J0812.4-3114 has been compared with our theoretical model under minimal cooling, we find that the thermal luminosity of RX J0812.4-3114 can be explained with low mass NS ($<1M_\odot$) under minimal cooling, which qualitatively agree with those in Ref.~\cite{Zhao2019}. However, to explain the upper limit of the high thermal luminosity of RX J0812.4-3114, other heating mechanism besides standard deep crustal heating should be considered. For example, the effect of X-ray burst may make the NS warm~\cite{Dohi2020,Ma2020,Meisel2019}, there are few works on the effect of X-ray burst on the quiescent luminosity of accreting NSs. Another possible mechanism is related to magnetic field, as RX J0812.4-3114 is a BeXRP which include a highly magnetised NS, the high magnetic field may effect the accretion and heating process compared with low-mass X-ray binaries which also may make the NS warm~\cite{Tsygankov2017,Gao2017,Potekhin2018}. It is also possible that the NS in RX J0812.4-3114 is still hot from supernova explosion as the previous work proposed~\cite{Zhao2019}. Furthermore, as mentioned in the paper, some physics input such as PBF neutrino emissivity and electron-ion bremsstrahlung neutrino emissivity are outdated, updating the code with modern microphysics input may also improve the quiescent luminosity and get more accurate results. We intend to tackle these interesting issues in a forthcoming work.

\begin{acknowledgements}
It is a great pleasure to acknowledge the very helpful remarks by the anonymous referee. 
This work has been supported by Xinjiang Science Fund under No. 2020D01C063, the National Natural Science Foundation of China under Nos. 11803026, 11473024, XinJiang University Science Fund, XinJiang Science Fund of 2017 Tianchi Program and the XinJiang Science Fund for Distinguished Young Scholars under No. QN2016YX0049. AD wishes to acknowledge the support from the Program of Interdisciplinary Theoretical \& Mathematical Sciences (iTHEMS) at RIKEN.
\end{acknowledgements}


\end{document}